\documentclass{jpp}
\usepackage{graphicx}
\usepackage{epstopdf, epsfig}
\usepackage{hyperref}
\usepackage{color}
\usepackage{mathtools}

\newcommand{\be}{\begin{equation}}
 \newcommand{\ee}{\end{equation}}

\shorttitle{Density functional theory for collisionless plasmas}
\shortauthor{G. Manfredi}

\title{Density functional theory for collisionless plasmas -- equivalence of fluid and kinetic approaches}

\author{Giovanni Manfredi\aff{1}
  \corresp{\email{giovanni.manfredi@ipcms.unistra.fr}} }
\affiliation{\aff{1} Universit\'e de Strasbourg, CNRS, Institut de Physique et Chimie des Mat\'eriaux de Strasbourg, UMR 7504, F-67000 Strasbourg, France
}

\begin{document}

\maketitle

\begin{abstract}
Density functional theory (DFT) is a powerful theoretical tool widely used in such diverse fields as computational condensed matter physics, atomic physics, and quantum chemistry. DFT establishes that a system of $N$ interacting electrons can be described uniquely by its single-particle density $n(\boldsymbol{r})$, instead of the $N$-body wave function, yielding an enormous gain in terms of computational speed and memory storage space. Here, we use time-dependent DFT  to show that a classical collisionless plasma can always, in principle, be described by a set of fluid equations for the single-particle density and current. The results of DFT guarantee that an exact closure relation, fully reproducing the Vlasov dynamics, necessarily exists, although it may be complicated (nonlocal in space  and time, for instance) and difficult to obtain in practice. This goes against the common wisdom in plasma physics that the Vlasov and fluid descriptions are mutually incompatible, with the latter inevitably missing some ``purely kinetic" effects.
\end{abstract}

\section{Introduction} \label{intro}
\subsection{Vlasov and fluid models}
It is  common knowledge among plasma physicists that a collisionless plasma should be described by a kinetic equation of the Vlasov type. Such an equation is  obtained formally from the $N$-body Coulomb problem through a BBGKY hierarchy procedure that amounts to neglecting two-body and higher-order correlations (``collisions"). The Vlasov equation describes the evolution of the distribution function $f(\boldsymbol{r}, \boldsymbol{v},t)$ in the six-dimensional (6D) phase space. Given the high dimensionality of the phase space, the solution of the coupled Vlasov-Maxwell or Vlasov-Poisson equations is a formidable computational task. In order to simplify this task, one of the most common strategies is to take velocity moments of the  distribution functions, leading to a system of fluid (or hydrodynamic) equations for such macroscopic quantities as the particle density $n(\boldsymbol{r},t)$ (zeroth-order moment of $f$), the current density $\boldsymbol{j}(\boldsymbol{r},t)$ (first-order moment), the pressure $P(\boldsymbol{r},t)$ (second-order moment), and so on.

The fluid models are obviously more tractable computationally compared to the original Vlasov equation, because all fluid quantities depend only on the space variables (plus time), and not on the velocity variables, which have been integrated away. Needless to say, the fluid models can never reproduce the full richness of the Vlasov approach, as some information is inevitably lost when performing the integrals in the velocity moment expansion. A blatant example is Landau damping, perhaps the most celebrated plasma effect, which is well present in Vlasov models, but totally absent in the fluid description.

But is it really so?  \citet{Hammett1990} already questioned this common belief in the early 1990s by showing that Landau damping can be reproduced (at least in the linear response regime) through a judicious choice of the fluid closure. The closure they proposed is local in Fourier space, but nonlocal in real space, so that it can be expressed in an integral form.

Another example of  exact correspondence between the fluid and the Vlasov approaches is provided by the water-bag model \citep{Bertrand1968}, which has been well studied since the 1960s. A  ``water-bag" distribution function is a special form of $f$ which is constant within a certain contour of the phase space and vanishing outside. By virtue of Liouville's theorem, this property is exactly preserved by the Vlasov equation. It can be shown that the two-fluid equations obtained from the water-bag distribution are  equivalent to the corresponding Vlasov equation. This equivalence is strict as long as the phase-space contour is convex, which precludes the formation of shocks.

Then, the question naturally arises   whether it is  always possible, at least in principle, to construct a closed system of fluid equations that are fully equivalent (and not just an approximation) to the original Vlasov model. Here, I will show  this is indeed possible under rather weak conditions. In other words, \emph{it is perfectly feasible to describe the full collisionless electron dynamics within the framework of fluid theory}. This is a rather bold conceptual step which should give fluid theory a new, more profound status. Nevertheless, not to overstate the preceding claim, I hasten to add this is a result of principle. It only proves that it is possible to close the system of fluid equations in an exact way, but does not offer any constructive method to derive such closure, although it does suggest some ways to improve our current closures.
Further, the closure relation will in general be nonlocal in space  and time and dependent on the initial phase-space distribution function (although the latter need not be a stationary state).

\subsection{Density functional theory}
Density-functional theory has had a huge impact on condensed-matter physics and quantum chemistry over the last few decades, since its onset in the mid 1960s. It was developed originally in two seminal papers by \citet{Hohenberg1964} and \citet{Kohn1965} to describe the ground state of a system of many electrons confined by an external potential $V(\boldsymbol{r})$. Hohenberg and Kohn showed that  the ground state can be expressed \emph{exactly} in terms of the one-body electron density $n(\boldsymbol{r})$, instead of the $N$-body wave function $\Psi(\boldsymbol{r}_1, \boldsymbol{r}_2, \dots \boldsymbol{r}_N)$,  which is  of course a huge simplification. They did so by establishing a one-to-one correspondence between the  density and the potential $n  \leftrightarrow V$. Since the ground-state wave function is fully determined by $V$ (if the ground state is not degenerate), this results in a one-to-one correspondence between the  density and the ground-state wave function $n  \leftrightarrow \Psi$, which is basically the content of the first Hohenberg-Kohn theorem (the second theorem states that the ground-state density can be derived by minimizing a certain energy functional). Shortly afterwards, \citet{Kohn1965} proposed a practical scheme to compute the ground-state density, known as the Kohn-Sham equations, which is still widely used today.

For instance, the Thomas-Fermi theory \citep{Fermi1927, Thomas1927} [for a modern account see \citep{Michta2015}] of the atomic electron gas can be viewed as an \textit{ante-litteram} version of modern DFT. In the Thomas-Fermi theory, the external potential is that of the atomic nucleus, $V(\boldsymbol{r})=-Ze/4\pi\epsilon_0 |\boldsymbol{r}|$. The total energy functional to be minimized is the sum of the external energy, the internal self-consistent Coulomb energy, and the kinetic energy: $E_{\rm TF}[n] = E_{\rm ext}[n] + E_{\rm int}[n] + T[n]$. These terms can be written respectively as:
\[
E_{\rm ext}[n] = e\int n(\boldsymbol{r})V(\boldsymbol{r}) d\boldsymbol{r},
\]
\[
E_{\rm int}[n] =  {1 \over 2}
\frac{e^2}{4 \pi \epsilon_0} \int\int  \frac{n(\boldsymbol{r}) n(\boldsymbol{r}')}
{|\boldsymbol{r}- \boldsymbol{r}'|}\, d\boldsymbol{r}\, d\boldsymbol{r}',
\]
and
\[
T[n] = {3 \over 10}\,(3\pi^2)^{2/3}\,{\hbar^2 \over {m}}\int n^{5/3} d\boldsymbol{r}.
\]
The total energy functional to be maximized is then: $F[n] = E_{\rm TF}[n] - \mu \int n d\boldsymbol{r}$, where $\mu$ is the chemical potential, which acts as a Lagrange multiplier to keep the total number of electrons fixed. By setting the functional derivative  of $F[n]$  equal to zero, one gets:  $\delta F[n]/\delta n - \mu =0$, which yields the standard Thomas-Fermi equation as found in textbooks:
\be
e V(\boldsymbol{r}) + \frac{e^2}{4 \pi \epsilon_0} \int \frac{n(\boldsymbol{r}') }{|\boldsymbol{r}'- \boldsymbol{r}'|}\, d\boldsymbol{r}' +
{\hbar^2 \over {2m}} \,(3\pi^2)^{2/3}\,n(\boldsymbol{r})^{2/3}  = \mu.
\ee
Of course, the Thomas-Fermi model is a very crude theory (it does not even predict the stability of atoms). But the point is that, by refining the energy functional $E[n]$ beyond the Thomas-Fermi approximation $E_{\rm TF}[n] $, one could in principle obtain an equation that predicts the \emph{exact} density $n(\boldsymbol{r})$ as given by the full $N$-body theory. \citet{Kohn1965} proved  that such exact energy functional exists, although it is not known explicitly and can only be approximately guessed.

Modern DFT was later generalized to the time-dependent case (TD-DFT) by \citet{Runge1984} [for reviews, see \citet{Marques2004, Ullrich2014}]\footnote{As of today, the original paper \citet{Runge1984} has been cited about 7000 times. Even a partial bibliography on TD-DFT is thus clearly out of the question here.}. These authors proved that the same initial quantum state evolving in two different confining potentials yields  different electron densities at all subsequent times  (more details about this statement will be given later). This establishes a one-to-one correspondence between the time-dependent potential and the density: $V(\boldsymbol{r},t)  \leftrightarrow n(\boldsymbol{r},t) $. Since the  wave function is fully determined by $V$, this fact translates into a one-to-one correspondence between the  time-dependent density and the time-dependent wave function: $n(\boldsymbol{r},t)  \leftrightarrow \Psi(\boldsymbol{r}_1 \dots  \boldsymbol{r}_N ,t) $. Again, this is a proof of principle, which does not furnish an explicit prescription how to construct such density. However, along the same lines as was done for the stationary case, one can obtain a set of time-dependent Kohn-Sham equations that yield the correct one-body density with great accuracy.

Apart from applications to condensed matter and atomic/molecular physics, TD-DFT is commonly used to describe the dynamics of strongly coupled plasmas in the so-called ``warm dense matter" regime [see, for instance, \citet{Magyar2016}]. This is a physical regime at the border between condensed matter and plasma physics,  where quantum and correlation effects cannot be neglected.

Here, the Runge-Gross theorem will be applied to collisionless plasmas described by the Vlasov-Poisson equations to show that it is always possible, in principle, to construct a fluid model that reproduces exactly the kinetic results.

\section{Runge-Gross theorem for collisionless plasmas} \label{RGtheorem}

In order to illustrate this result, we consider the simple case of a single-species plasma confined in an external potential in one dimension (1D), governed by the Vlasov-Poisson equations:
\begin{eqnarray}
\frac{\partial{f}}{\partial{t}} &+& v\frac{\partial{f}}{\partial{x}} - \frac{q}{m} \left(\frac{\partial \phi}{\partial{x}} + \frac{\partial V}{\partial{x}}\right) \frac{\partial{f}}{\partial{v}}  = 0, \label{eq:vlasov}\\
\frac{\partial^2 \phi}{\partial x^2}  &=& - {q \over \epsilon_0} \int_{-\infty}^{+\infty} f(x,v,t)\,dv =  -{q  \over \epsilon_0}  n(x,t),
\label{eq:poisson}
\end{eqnarray}
where $q$ and $m$ are respectively the charge and mass of the particles, $V(x,t)$ is the external confining potential, and $\phi(x,t)$ is the self-consistent electrostatic potential.

The Runge-Gross (RG) theorem \citep{Runge1984} can be stated as follows, using the language of 1D plasmas relevant to the present work:\\
{\it Theorem}.---
For every external potential $V(x, t)$ which can be expanded into a Taylor series with respect to the time coordinate around $t=0$, a
map $V(x,t) \to n(x,t)$ is defined by solving the Vlasov-Poisson equations with an initial state $f(x,v,t=0)=f_0(x,v)$, and calculating the corresponding distribution function $f(x,v,t)$ and particle density $n=\int f dv$ at a subsequent time $t$. This map can be inverted up to an additive merely time-dependent function, i.e. $n(x,t) \to V(x,t) + c(t)$.

The theorem establishes a one-to-one correspondence between $n(x,t)$ and $V(x,t)$. Since $V$ determines the evolution of $f$, it proves that the full kinetic evolution of the system can be encoded in the density (without necessarily computing the distribution function).
We note that this correspondence is over the full space-time evolutions of $n$ and $V$, in a form depending also
on the initial distribution function $f_0(x,v)$.

 {\it Proof}.--- To prove the RG theorem is equivalent to demonstrating that the particle densities $n(x, t)$ and $n'(x, t)$, evolving from the same initial state $f_0(x,v)$ under the influence of the two external  potentials $V(x, t)$ and $V'(x, t)$, are always different provided that the two potentials differ by more than a function of time, ie:
 \be
 V(x, t) - V'(x, t) \neq c(t)
 \label{eq:potdiff}
 \ee
 We first prove that the corresponding current densities $j(x,t)$ and $j'(x,t)$ are different, then we extend the proof to the densities: in other words, we first prove that $j \to V$, and then that $n \to j$.

 Let us write the equation of motion for the current density, by taking the first velocity moment of the Vlasov equation (\ref{eq:vlasov}):
 \be
 \frac{\partial j}{\partial t} = -  {2 \over m} \frac{\partial T}{\partial x} - n {q\over m} \frac{\partial \phi}{\partial x} -n {q\over m}\,\frac{\partial V}{\partial x},
 \label{eq:currdens}
 \ee
 where
 \(
 T(x,t) = {m \over 2} \int f v^2dv
 \)
 is the kinetic energy density. (In the definitions of the velocity moments, the velocity integrals will  be assumed to extend to $\pm \infty$ and the distribution function to decay sufficiently rapidly for those moments to exist and be finite). 
 An  equation similar to (\ref{eq:currdens}) can be written for $j'$. Since the initial state $f_0(x,v)$ is the same for both cases, all terms on the right-hand side of (\ref{eq:currdens}) are identical at the initial time $t=0$, except for the last one (external potential).
 Taking the difference between equations for $j$ and $j'$ at  $t=0$, one obtains
  \be
 {\frac{\partial (j-j')}{\partial t} \vline}_{\,t=0} = - n{q\over m} {\frac{\partial (V-V')}{\partial x}\vline}_{\,t=0} .
 \label{eq:currdiff}
 \ee
 If the potentials differ at $t=0$, then the right-hand side of (\ref{eq:currdiff}) will be different from zero and thus $j(x,t)$ and $j'(xt,t)$ will become different infinitesimally later than the initial time. In this case, the first part of the proof (ie, $j \to V$) is almost trivial.

Next, we consider the case where $V(x,0)=V'(x,0)$, but their first time derivatives differ, ie: $\partial_t [V(x,t)-V'(x,t)]_{t=0} \neq c(t)$. In this case, (\ref{eq:currdiff}) cannot be used to predict the divergence of $j$ and $j'$ for $t>0$. In order to prove the RG theorem, we need to differentiate (\ref{eq:currdens}) with respect to time. Taking the second-order velocity moment of the Vlasov equation (\ref{eq:vlasov}), we get:
\be
\frac{\partial T}{\partial t} =  -  \frac{\partial Q}{\partial x} +q j \frac{\partial U }{\partial x},
\label{eq:evol-T}
\ee
where $Q(x,t)={m \over 2}\int v^3 f dv$ is the kinetic energy flux and we defined $U\equiv V+\phi$.  Inserting (\ref{eq:evol-T}) into the time derivative of (\ref{eq:currdens}), we obtain:

\begin{eqnarray}
\frac{\partial^2 j}{\partial t^2} &= & {2 \over m}\,\frac{\partial }{\partial x} \left(\frac{\partial Q}{\partial x}- qj\frac{\partial U}{\partial x} \right) -
{q \over m}  \left[\frac{\partial U}{\partial x}\frac{\partial n}{\partial t} +
n \frac{\partial }{\partial x} \frac{\partial U}{\partial t} \right]  \nonumber \\
&= & {2 \over m}\,\frac{\partial }{\partial x} \left(\frac{\partial Q}{\partial x}- qj\frac{\partial U}{\partial x} \right) + {q \over m} \left[\frac{\partial U}{\partial x}\frac{\partial j}{\partial x} +
n \frac{\partial }{\partial x} \left(G \star \frac{\partial j}{\partial x} \right) -
n \frac{\partial }{\partial x} \frac{\partial V}{\partial t}
\right]
 \label{eq:currdens2}
\end{eqnarray}
To obtain  (\ref{eq:currdens2}), we made use of the continuity equation $\partial_t n=-\partial_x j$ and of the fact that the self-consistent potential $\phi[n]$ is a  linear functional of the density
\be
\phi[n] = \int G(x,x')\,n(x',t) \,dx' \equiv G \star n,
\label{eq:poissoninteg}
\ee
where $G(x,x')$ is the integral kernel (Green function) of the  interaction, which also depends on the boundary conditions and on the dimensionality of the system. By choosing the Coulomb kernel  $G_{\rm Coul}(x,x') \equiv -(q/\epsilon_0)\,|x-x' |$,  then \eqref{eq:poissoninteg} is just the integral form of the 1D Poisson equation \eqref{eq:poisson}.
Taking the time derivative and using once again the continuity equation, one obtains:
\be
\frac{\partial \phi}{\partial t} = - \int G(x,x')\,\frac{\partial j}{\partial x} (x',t)\,dx' = -G \star \frac{\partial j}{\partial x} ,
\ee
which was used to obtain the last line of \eqref{eq:currdens2}.
 However, it is important to stress that, for the proof of the RG theorem, $G$ is not restricted to be the kernel that corresponds to the Poisson equation. It is only required that the potential $\phi$ may be written as a functional of the density, as in  \eqref{eq:poissoninteg}.

Then, we can write the same expression as  \eqref{eq:currdens2} for $j'$ and note that, if we assume $V(x,0)=V'(x,0)$,  all term on the right-hand side are identical at the initial time. Taking the difference between the two equations, one gets at $t=0$:
\be
{\frac{\partial^2 (j-j')}{\partial t^2} \vline}_{\,t=0} = -{q \over m} n\, \frac{\partial }{\partial x} {\left(\frac{\partial}{\partial t}  (V-V') \right)\vline}_{\,t=0} \,,
\label{eq:currdiff2}
\ee
Now, if
\(
\partial_t (V-V') \vline_{\,t=0}  \neq c(t) ,
\)
then the two currents $j$ and $j'$ start diverging right after the initial time, again in accordance with the RG theorem.

The above reasoning can be extended to the case where $V$ and $V'$ differ at a higher-order time derivative. One obtains:
\be
{\frac{\partial^{k+1} (j-j')}{\partial t^{k+1}} \vline}_{\,t=0} = -{q \over m} n \frac{\partial }{\partial x} {\left(\frac{\partial^k}{\partial t^k}  (V-V') \right)\vline}_{\,t=0} \,.
\label{eq:currdiffk}
\ee
Therefore, the currents $j$ and $j'$ diverge if there exists some nonnegative integer $k$ such that
\be
w_k(x) \equiv \frac{\partial^k}{\partial t^k} {\left[ V(x,t)-V'(x,t) \right]}_{t=0} \neq \rm const.,
\label{eq:wk}
\ee
which is a consequence of the assumption (\ref{eq:potdiff}) and  the requirement that the external potential is expandable in a Taylor series around $t=0$.

This completes the first part of the proof, namely that $j \to V$. Next, we turn to the densities and consider the continuity equation:
\be
\frac{\partial n}{\partial t} = - \frac{\partial j}{\partial x} .
\label{eq:continuity}
\ee
Taking the $(k+1)$-th derivative of the continuity equation and using (\ref{eq:currdiffk}), we obtain
\be
{\frac{\partial^{k+2} (n-n')}{\partial t^{k+2}} \vline}_{\,t=0} = \frac{q}{m} {\frac{\partial }{\partial x} \left[ n \frac{\partial }{\partial x} \left(\frac{\partial^k}{\partial t^k}  (V-V') \right) \right]\vline}_{\,t=0} \equiv
 \frac{q}{m} {\frac{\partial }{\partial x} \left( n_0 \frac{\partial w_k}{\partial x} \right)}  ,
\label{eq:ndiffk}
\ee
where $n_0(x)=n(x,t=0)$.
All we need to prove is that the right-hand side of (\ref{eq:ndiffk}) cannot vanish if (\ref{eq:wk}) holds. Let us consider the integral
\be
\int_{-\infty}^{+\infty} n_0\left( \frac{\partial w_k }{\partial x} \right)^2 dx = - \int_{-\infty}^{+\infty} w_k \frac{\partial }{\partial x}\left( n_0 \frac{\partial  w_k}{\partial x} \right) dx + {1 \over 2} n_0 { \frac{\partial w_k^2}{\partial x}\vline}_{-\infty}^{+\infty} \, .
\label{eq:lemma}
\ee
Assuming that the density falls sufficiently fast at infinity,  the boundary term in (\ref{eq:lemma}) vanishes. Then, if (\ref{eq:wk}) holds, the left-hand side of (\ref{eq:lemma}) cannot vanish, and hence
\[
\frac{\partial }{\partial x} \left( n_0 \frac{\partial w_k}{\partial x} \right)
\]
cannot vanish everywhere.
This complete the second part of the proof that the density determines the current, ie, $n \to j$.
For some limitations and mathematical subtleties on the validity of the RG theorem, see \citet{Dhara1987}.

\begin{figure}
  \centerline{\includegraphics[scale=0.3]{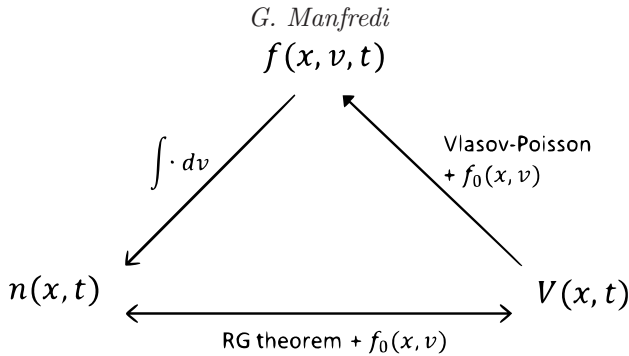}}
  \caption{Schematics of the TD-DFT approach. The RG theorem proves the equivalence between $n$ and $V$. But $V$ determines $f$ through the Vlasov-Poisson equations and $n$ is obtained from $f$ by velocity-space integration. Thus, the counterclockwise arrows show that $n \leftrightarrow f$, ie, the full Vlasov dynamics can be encoded in the evolution of the density. }
\label{fig:scheme}
\end{figure}

All in all, we have proven that $n \to V$, and since the potential $V$ trivially determines $n$ through the solution of the Vlasov-Poisson equations, one can finally deduce that the map between $n$ and $V$ is invertible, i.e. $n \leftrightarrow V$. This is summarized in the diagram of Fig. \ref{fig:scheme}.
We stress that this map is understood as being over the full space-time evolutions of $n$ and $V$, and also depends
on the initial distribution function $f_0(x,v)$.

\section{Consequences of the Runge-Gross theorem}\label{sed:conseq}
\subsection{Fluid equations}
By virtue of the RG theorem, one can represent the full kinetic dynamics in terms of the particle density $n$, and the current as an auxiliary variable. This is achieved through the momentum conservation and continuity equations (\ref{eq:currdens}) and (\ref{eq:continuity}), which we reproduce here for clarity:
\begin{eqnarray}
\frac{\partial n}{\partial t} &+&  \frac{\partial j}{\partial x}  = 0, \label{eq:fluid-cont}\\
 \frac{\partial j}{\partial t}  &=& -  {2 \over m}\, \frac{\partial T}{\partial x} - n {q\over m} \left( \frac{\partial \phi}{\partial x} + \,\frac{\partial V}{\partial x}\right) .
\label{eq:fluid-mom}
\end{eqnarray}
The point is that,  given the initial data  $f_0(x,v)$, we can express every term in the above equations as a functional of the density or the current (which is itself determined by $n$). This is already the case for the self-consistent potential $\phi[n]$, because it is a solution of Poisson's equation. The RG theorem implies that also the kinetic energy density can be expressed as  $T[n]$. This is what is usually called a ``closure" in fluid theory. The RG theorem provides no explicit recipe to obtain this functional, and finding accurate expressions is all the difficulty of DFT. First, we  separate the centre-of-mass kinetic energy and the pressure:
\be
T[n] = {m \over 2} {j^2 \over n} + {1 \over 2}P[n].
\label{eq:Tn}
\ee
 In the first term on the right-hand side of \eqref{eq:Tn}, $j$ is determined by the evolution of $n$ according to the continuity equation \eqref{eq:fluid-cont}.
As to the second term, we know, from the RG theorem, that there exists a functional $P_{\rm exact}[n]$ that makes the fluid equations (\ref{eq:fluid-cont})-(\ref{eq:fluid-mom}) exactly equivalent to the Vlasov equations.
 In contrast to $\phi[n]$ (which is an instantaneous functional of the density), the functional $P_{\rm exact}[n]$ will generally be nonlocal in space and time and dependent on the initial datum $f_0(x,v)$.
Note, however, that $f_0(x,v)$ need not be a stationary state.
The challenge is then to find an accurate approximation for $P_{\rm exact}[n]$. Of course, many closure relations have been proposed in plasma physics, notably polytropic equations of state of the type $P[n] \propto n^\gamma$,  which are valid for specific types of equilibrium states.

\subsection{Fluid closures}
Here, we will revisit two closures which, as mentioned in the Introduction, challenge the common belief that the fluid equation are necessarily less complete than the corresponding kinetic model.
In addition to these classical closures, we should also mention that a nonlocal fluid closure that is exact within the linear response has recently been presented by \citet{Moldabekov2018} for the case of a quantum plasma.

\subsubsection{Water-bag model}
The water-bag model \citep{Bertrand1968,Feix2005} is based on a special form of the distribution function, which is constant ($f=\mathcal{A}$) within a certain closed contour in the phase space,  and $f=0$  outside. This property is preserved during the Vlasov evolution and only the shape of the contour is modified. As long as the contour can be defined by two single-valued curves $v_{+}(x,t)$ and  $v_{-}(x,t)$, for the maximum and minimum velocity respectively, the Vlasov equation is strictly equivalent to the two equations for the contour evolution:
\begin{eqnarray}
\frac{\partial v_+}{\partial t} +v_+ \frac{\partial v_+}{\partial x}   &=& -{q \over m} \frac{\partial U}{\partial x} , \label{eq:vplus} \\
\frac{\partial v_-}{\partial t} +v_- \frac{\partial v_-}{\partial x}   &=& -{q \over m} \frac{\partial U}{\partial x} ,
\label{eq:vminus}
\end{eqnarray}
where $U=\phi + V$ includes both the self-consistent and the external potentials.

With this prescription, it is easy to compute the particle and current densities:
\be
n= \int_{v_-}^{v_+} f dv = {\mathcal{A} \over 2} (v_+ + v_-);\,\,\,\,\,\,\,
j = \int_{v_-}^{v_+} vf  dv = {\mathcal{A} \over 2} (v_+^2 -  v_-^2).
\ee
The sum of equations (\ref{eq:vplus}) and  (\ref{eq:vminus}) yields the continuity equation. Their difference yields the following momentum conservation equation
\be
\frac{\partial j}{\partial t}   + \frac{\partial }{\partial x} \left({j^2 \over n}\right) = -{q \over m} \frac{\partial U}{\partial x} - {1 \over m}\frac{\partial P}{\partial x},
\label{eq:wbcurr}
\ee
where
\be
P[n] = {m \over {12 \mathcal{A}^2}}\, n^3.
\label{eq:wbpress}
\ee
Equations (\ref{eq:wbcurr})-(\ref{eq:wbpress}) are {\em exactly} equivalent to the Vlasov equation, as long as the contours $v_\pm$ remain single-valued. This is therefore a special case where the functional $P_{\rm exact}[n]$ is perfectly known. Note that the closure (\ref{eq:wbpress}) corresponds to an adiabatic equation of state. The strict equivalence is broken when the velocity contours become multivalued: in that case, although (\ref{eq:vplus})-(\ref{eq:vminus}) still hold, the fluid equations (\ref{eq:wbcurr})-(\ref{eq:wbpress}) are no longer exact.

\subsubsection{Landau fluid models}
In the early 1990s, Hammett and coworkers tried to include (linear) Landau damping in a fluid model \citep{Hammett1990}. They considered a 1D electron plasma in an infinite medium, with homogeneous immobile ions. As \citet{Hammett1992} showed, common fluid closures where the pressure is an algebraic function of the density or where the $l$-th order moment is set to zero, fail to reproduce Landau damping, because all coefficients in the corresponding dispersion relation are real, yielding both damped and unstable modes.

Nevertheless, it is possible to include  damping in the fluid equations, if the closure relation depends on the {\em gradients} of the lower-order moments. The simplest example is Fick's law, $j=-D\partial_x n$, which yields the diffusion equation for $n$. Extending to the second-order moment \citet{Hammett1992}, one can approximate the pressure as:
\be
P = n k_B T_0 - \mu_k m n {{\partial( j/n)}\over {\partial x}},
\label{eq:pclosure}
\ee
where $T_0$ is a constant temperature, $k_B$ is Boltzmann's constant, and $\mu_k$ is a parameter  that plays the role of a viscosity. The simple prescription (\ref{eq:pclosure}) already introduces an imaginary term in the  dispersion relation and therefore can lead to damping. If, in addition, the parameter $\mu_k$ is taken to be a function of the wave number $k$, then the dispersion relation can be tailored to reproduce the kinetic Landau damping to some accuracy.
For instance, \citet{Hammett1992}  took $\mu_k= \sqrt{\pi/2}\,v_{th}/|k|$ (where $v_{th}$ is the thermal speed), which yields for the imaginary part of the frequency: ${\rm Im}\, \omega = -\sqrt{\pi/8}\,|k| v_{th}$.
In this case, the closure relation in real space becomes an {\em integral} one.

Better approximations can be obtained by carrying out the expansion to higher-order moments (a three-moment fluid system is required to conserve energy, for example). More generally, it can be shown  \citep{Hammett1990}  that one can always construct  an equilibrium distribution that exactly yields the $l$-moment closure. So, at least at the level of the linear response, we know that some kinetic evolutions (including Landau damping) can be exactly represented by a fluid model. For more recent work on Landau fluid models, see for instance \citet{Hunana2018}.
This is in line with the formal result of the RG theorem. Also note that,  in standard DFT, so-called generalized gradient approximations (GGA) are commonly used to express the kinetic energy functional \citep{Perdew1992}.

A challenge with viscosity-based closures is that they break the entropy conservation of the Vlasov equation. This is not a big problem for the linear response considered in the above-cited papers, as the linearized Vlasov equation does not preserve entropy anyway.
For the full nonlinear response, the equivalence between a dissipative fluid model and an entropy-preserving Vlasov equation should involve a connection which is nonlocal in time.

\subsection{Equivalence to a noninteracting system}\label{subsec:equivalence}
In the Kohn-Sham equations of DFT \citep{Kohn1965}, the kinetic energy is approximated by considering an equivalent noninteracting system with the same density $n$ as the original interacting one. Here, we can also show a  similar equivalence, as was done by  \citet{Leeuwen1999} and \citet{Vignale2004} for TD-DFT.
Indeed, in our proof of the RG theorem,  we only used the fact that the interaction potential $\phi[n]$ is a functional of the density, as in  \eqref{eq:poissoninteg}. But the Green function $G(x,x')$ needs not be the one of the Poisson equation for the proof to be valid. For instance, for a contact interaction with coupling constant $g$, one has $G(x,x') = g\delta(x-x')$, so that $\phi = gn$, and the proof of the RG theorem is still valid.

\begin{figure}
  \centerline{\includegraphics[scale=0.3]{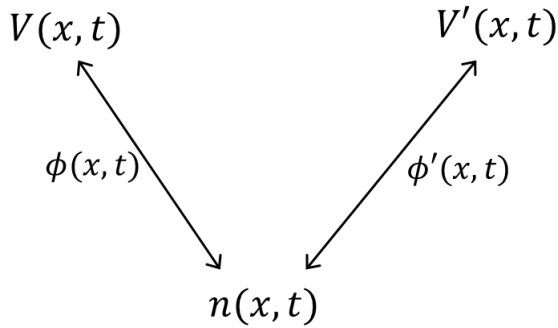}}
  \caption{There is a one-to-one correspondence between the density $n(x,t)$ and the external potentials  $V$ and $V'$, for different interactions $\phi$ and $\phi'$.  }
\label{fig:scheme2}
\end{figure}

Therefore, the result
\[ n \xleftrightarrow[]{{\,\,\,\phi[n]}\,\,\,} V
\]
is independent of the specific choice of the potential. We could have carried out the same derivation for another interaction potential $\phi'[n]$ and demonstrated the equivalence
\[
n \xleftrightarrow[]{\,\,{\phi'[n]}\,\,} V' .
\]
This is summarized in Fig. \ref{fig:scheme2}.

But one could also take $\phi'=0$ (noninteracting particles) and the theorem would still hold. This means that we can recover the correct time-dependent density $n(x,t)$ of the interacting Coulomb system with external potential $V$ by solving the fluid equations for a noninteracting system with external potential $V'$ (unknown {\it a priori}, but which could be guessed or approximated on physical grounds). This approach might reveal itself fruitful for the task of approximating the kinetic energy functional $T[n]$ in the fluid equation (\ref{eq:currdens}).
Incidentally, the equivalence between the interacting and noninteracting systems provides a justification for the fact that closures based on the  equation of state of an ideal (ie, noninteracting) gas, $P = n k_B T$, work relatively well in many situations.

\section{Conclusions}\label{sec:conclusions}
The purpose of this paper was primarily pedagogical. We showed that the results of time-dependent density functional theory (widely used in condensed-matter and high energy-density physics) can be applied to collisionless (Vlasov) plasmas. The main result is that the description of such a plasma can be done purely in terms of the particle density, instead of the phase-space distribution function. We presented a proof of this result (Runge-Gross theorem) for the simple case of an electron plasma confined in a 1D potential well, which can be extended in a straightforward way to the 3D case. Hence, the full plasma dynamics (including typical ``kinetic" effects, such as Landau damping) can be exactly described by a set of fluid equations, endowed with an appropriate closure for the kinetic energy density $T[n]$.
Such a closure is generally nonlocal in space and time and dependent on the initial phase-space distribution function.

Although it does not provide an explicit recipe to construct $T[n]$, which must therefore be approximated, the RG theorem has nevertheless a considerable theoretical reach. It puts  fluid theories on a solid conceptual basis as being capable of accounting, at least in principle, for all plasma effects, even those usually considered as ``purely kinetic".
This result may also be useful in practice when searching for better expressions of the closure relations, for instance for improving the Landau fluid closures. The closures first proposed by \citet{Hammett1990} are capable of recovering damping (to some approximation) because they depend on the {\em gradients} of the lower-order moments. In addition, to  properly model Landau damping, one has to resort to nonlocal (integral) expressions. However, these closures remain local in time (ie, adiabatic). A generalization to closure relations that depend on the past history of the system (non-adiabatic) may improve the reach and accuracy of this approach.

It should also be mentioned that, in standard TD-DFT, it is possible to include damping effects in a local way if the theory is formulated in terms of the {\em current density} instead of the particle density \citep{VignaleKohn1996,Vignale1997} (this approach is referred to as current-DFT). Such developments may also be useful for collisionless plasma applications.
Likewise, the considerations of Sec. \ref{subsec:equivalence} may suggest new closures based on an equivalent noninteracting system.
Finally, improved closures may also be determined empirically by systematic computer simulations of the kinetic model, from which the relationship between pressure and density is extracted numerically.


\bibliographystyle{jpp}

\bibliography{bibliojpp}

\end{document}